\documentclass[sigconf]{acmart}
\usepackage{balance}

\AtBeginDocument{%
  \providecommand\BibTeX{{%
    \normalfont B\kern-0.5em{\scshape i\kern-0.25em b}\kern-0.8em\TeX}}}

\setcopyright{rightsretained}
\copyrightyear{2021}
\acmYear{2021}
\acmBooktitle{Contribution to the CHI'21 Workshop on Emergent Interaction: Complexity, Dynamics, and Enaction in HCI}
\begin{document}

%%
%% The "title" command has an optional parameter,
%% allowing the author to define a "short title" to be used in page headers.
\title{Simulating Social Acceptability With Agent-based Modeling}

%%
%% The "author" command and its associated commands are used to define
%% the authors and their affiliations.
%% Of note is the shared affiliation of the first two authors, and the
%% "authornote" and "authornotemark" commands
%% used to denote shared contribution to the research.
\author{Alarith Uhde}
\email{alarith.uhde@uni-siegen.de}
\orcid{0000-0003-3877-5453}
\affiliation{%
  \institution{Siegen University}
  \streetaddress{Kohlbettstraße 15}
  \city{Siegen}
  \country{Germany}
  \postcode{57072}
}
\author{Marc Hassenzahl}
\email{marc.hassenzahl@uni-siegen.de}
\orcid{0000-0001-9798-1762}
\affiliation{%
  \institution{Siegen University}
  \streetaddress{Kohlbettstraße 15}
  \city{Siegen}
  \country{Germany}
  \postcode{57072}
}

%%
%% By default, the full list of authors will be used in the page
%% headers. Often, this list is too long, and will overlap
%% other information printed in the page headers. This command allows
%% the author to define a more concise list
%% of authors' names for this purpose.
%\renewcommand{\shortauthors}{Trovato and Tobin, et al.}

%%
%% The abstract is a short summary of the work to be presented in the
%% article.
\begin{abstract}
Social acceptability is an important consideration for HCI designers who develop technologies for social contexts. However, the current theoretical foundations of social acceptability research do not account for the complex interactions among the actors in social situations and the specific role of technology. In order to improve the understanding of how context shapes and is shaped by situated technology interactions, we suggest to reframe the social space as a dynamic bundle of social practices and explore it with simulation studies using agent-based modeling. We outline possible research directions that focus on specific interactions among practices as well as regularities in emerging patterns.
\end{abstract}

%%
%% The code below is generated by the tool at http://dl.acm.org/ccs.cfm.
%% Please copy and paste the code instead of the example below.
%%


%%
%% Keywords. The author(s) should pick words that accurately describe
%% the work being presented. Separate the keywords with commas.

%% A "teaser" image appears between the author and affiliation
%% information and the body of the document, and typically spans the
%% page.
%\begin{teaserfigure}
  %\includegraphics[width=\textwidth]{sampleteaser}
  %\caption{Seattle Mariners at Spring Training, 2010.}
  %\Description{Enjoying the baseball game from the third-base
  %seats. Ichiro Suzuki preparing to bat.}
  %\label{fig:teaser}
%\end{teaserfigure}

%%
%% This command processes the author and affiliation and title
%% information and builds the first part of the formatted document.
\maketitle

\section{Introduction}

Social acceptability is a long-standing theme of research in HCI~\cite{Koelle2020, Venkatesh2000}. It is concerned with the question how people decide whether or not to use a certain technology in a social context, and how technology can be designed so that it ``fits'' with social settings. Despite the importance of the issue, theoretical progress has been slow and the two most widely applied theories both have their specific drawbacks. On the one hand, the ``Technology Acceptance Model'' (TAM) was originally created for workplace settings with the goal to reduce the ``unwillingness'' of workers to use a technology~\cite{Davis1989, Venkatesh2000, Venkatesh2008}. The social factors used in the model to describe social acceptability include the degree to which a user thinks that the management and/or peers approve of the technology interaction, whether it is associated with a high status, and with a notion of compliance. This framing may still be suitable for efficiency-oriented work settings, but is arguably less applicable for other domains. On the other hand, Goffman's dramaturgical model of social interactions is focused on how people try to manage the impression they make on others. It is concerned with intentional and unintentional signals people send and perceive about each other in public settings, as well as their efforts to leave a positive impression. It can be applied to a broader range of social settings than the TAM, but it provides no specific insights on technology and its design - which was also not its focus~\cite{Goffman1959}.

% Typically framed from either the user or the ``observer'' perspective (or spectator, bystander, you name it), but group-level processes have not been studied so far. Also: interaction vs. context distinction, no ``shaping'' of context.

In addition, more recent empirical studies typically concentrate on the perspective of either the user or the ``observer'' (or spectator, bystander, audience)~\cite{Alallah2018, Dalsgaard2008, Koelle2020, Reeves2005}. These roles are reflective of a rather static view on the social context, with an active user who is interacting with a technology, and other people looking at him or her. Realistically, however, people outside of the lab are engaged in their own activities and we think that these activities, as well as their relationship or compatibility with each other, are crucial to better understand social acceptability (see also~\cite{Dourish2004, Reckwitz2002, Uhde2021b}). In other words, we think that the key to understand social context and its relevance for technology interaction does not lie in a better understanding of the involved individuals or locations as such. Instead, we need to consider what everyone is doing in that setting, and how a technology interaction relates to all the other situated activities~\cite{Dourish2004, Uhde2021b}. This results in a more dynamic perspective on technology interactions in a given context and allows us to not only understand the influence of the surroundings on the interaction, but also the active role this interaction plays in shaping the context. Such a decentered view on the problem of social acceptability allows us to simultaneously consider the activities of everyone involved and the specific ways these activities relate to each other.

% Einwurf: Social Practice Theory

Further support for this perspective and the importance of activities for shaping social context comes from social practice theory (e.g.,~\cite{Reckwitz2002, Shove2012}). Here, interactions (which are called ``social practices''), essentially consist of three types of components: Materials, competences, and meanings~\cite{Shove2012}. A specific practice can be performed if its necessary components come together, and through this performance it becomes a defining element of the situation~\cite{Reckwitz2002}. In fact, in social practice theory these practices represent the fundamental unit of social analysis. As an example, the practice of ``making coffee'' can be performed if a coffee machine (material), the ability to use it (competence), and the wish to drink coffee (meaning) come together (e.g.,~\cite{Klapperich2019b, Klapperich2020}). If the performance of this practice of making (and later drinking) coffee is joined by other practice performances such as reading a newspaper and eating a croissant, they can collectively form a ``breakfast'' context~\cite{Shove2009}. In this context, other breakfast practices also become more appropriate and incompatible ones (e.g., singing a lullaby) become less appropriate. In this way, social practice theory provides a bottom-up perspective about how contexts are shaped and does not depend on global factors (such as the clock time in this case or location categories).

% For the interaction of practices, HCI research was so far limited.
% 1. Lawo and friends: Complexes with SNA to identify e.g., central components and structural holes
% But: Bound by common material, which is not applicable in social acceptability research
% Also: Static (no time component)

A few previous studies provide more details about the relationships between practices~\cite{Higginson2015, Higginson2016, Lawo2020}. Specifically, they have used social network analysis~\cite{Otte2002} to visualize networks of practices that share common elements. For example, domestic food practices often revolve around certain material (food) and meanings (health or joy)~\cite{Lawo2020}. This approach proved useful to identify central components within complex networks of practices that share a common theme as well as structural holes that could be opportunities for innovation.

However, in the case of social acceptability, the central concern is not which elements are shared by many practices. Instead, separate practices with different components are simply co-located, and this sometimes fits well and sometimes not. The central question here is how this ``fit'' emerges. As a consequence, the types of relations between the practices need to be understood differently. Overall, they can support or conflict with each other, but how this looks like in detail has not been studied yet. For example, imagine you are trying to read a book while the person next to you is playing loud music. In this case, the relationship between the two practices could be understood as a conflict between the competence to concentrate on the book and the material (soundwaves) of the music practice, both of which are typically not considered as shared elements of the two practices. As a consequence of this conflict, noisy practices are typically not considered acceptable in reading spaces (such as libraries).

% Practice theory is helpful, but not specific enough.

We think that without more appropriate, dynamic models of interactions between practices in co-located settings, theoretical progress about social acceptability will remain limited. There are a number of research questions that require a broader perspective which cannot be covered with traditional methodology. Some of these require a simultaneous consideration of individual and group level processes. These questions include the following:

\begin{itemize}
 \item{How do co-located practices self-organize and emerge into a shared consensus about the type of context?}
 \item{What happens if no consensus can be reached (and what are the conditions for that)?}
 \item{How exactly do the components of practices (material, competence, meaning) relate to the components of surrounding practices?}
 \item{How does the context influence the selection of a specific practice among several options?}
 \item{How do individual practices contribute to changing a context into a different one? What does ``different'' mean here?}
 \item{How does the initial constellation of practices influence the further development of the context?}
 \item{How does the density of practices (``crowdedness'') influence social acceptability?}
 \item{What overarching patterns of acceptability can we observe across different contexts? To what degree are contexts similar and what are differences? Can these differences be described in a systematic way?}
 \item{How can we account for context in technology design? Where are the limits of compatibility across different contexts?}
\end{itemize}

As indicated above, in order to answer these questions we think that we need a different approach than what traditional methods used in social acceptability research can deliver. Experimental lab or field studies are useful for analyzing low numbers of factors if no other ethical or practical restrictions apply, but they are of limited value in cases (like this one) with multidirectional causality~\cite{VanBerkel2020}. Qualitative methods can provide rich insights into specific contexts and subjective experiences, but they do not provide the flexibility to study and experiment with different settings and have restrictions when fictional scenarios are studied. In order to complement and overcome some of these issues, we think that simulation studies, particularly agent-based modeling (ABM), can be useful here.

\section{Social Acceptability and Agent-based Modeling}

Agent-based modeling (ABM) is a method to study how interactions between individual ``agents'' on a local level lead to emergent patterns on a group level. In our case, the social practices and people (agents) interact locally which leads to a consensus about the context and socially acceptable behavior in the group (emerging pattern). ABMs have been used in a variety of fields and classic examples include the study of segregation in cities~\cite{Schelling1971}, bird flocking behavior~\cite{Reynolds1987}, and Conway's game of life~\cite{Gardner1970}. In computational social science, ABMs have been used before to simulate the evolution of social practices over time~\cite{Holtz2014} and the dynamics of social practices of energy consumption in households~\cite{Narasimhan2017}.

ABM is a simulation method, meaning that researchers experiment with virtual agents that interact with each other in a simulated scenario. Unlike research methods based on inductive and deductive reasoning, the insights ABM can provide are understood as generative~\cite{Axelrod1997}, which combines some aspects of both and is in a way comparable to some design research methods. The programmatic approach requires researchers to make their assumptions about rules of interaction explicit (like deductive approaches), which then facilitates critical discussion and enforces high specificity about the components of the model. The model can then be used to generate data and experiment with the model, which allows for inductive reasoning.

ABM has some advantages that we would like to point out briefly. First, ethical restrictions do not apply to ABMs the same way as they apply to other methods, e.g., simulating ``shameful'' behavior in certain social contexts is less problematic than experimenting with it in the wild. Second, ABMs allow for replication, to freely manipulate the environment and to study social phenomena at different scales, which can be used to increase confidence in the model~\cite{Grimm2005}. And third, one important advantage of ABMs is that they provide detailed process data about the development of a phenomenon over time, rather than only summative, retrospective data~\cite{Wilensky2015}. An additional important note is that ABMs do not aim to be completely realistic reproductions of reality. Instead, they help to emphasize and experiment with specific aspects of reality and explore behavior. Similar to other methods, quality criteria exist for verification and validation of ABMs~\cite{Rand2011, Wilensky2015}.

\begin{figure}[h]
  \centering
  \includegraphics[width=\columnwidth]{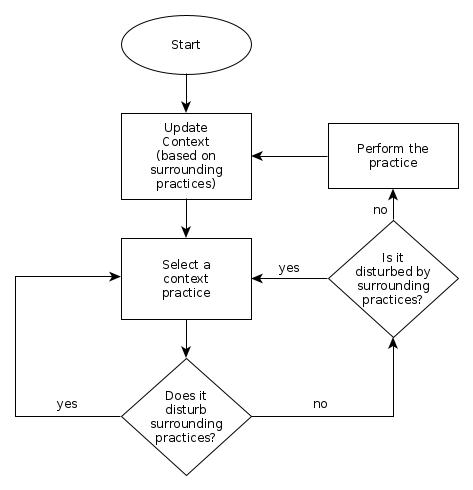}
  \caption{A basic example for a decision-making process of a human agent in a social context.}
  \Description{A flowchart depicting the decisions of a human agent to perform a certain social practice in a social context. The steps are described in the main text.}
  \label{flowchart}
\end{figure}

Figure~\ref{flowchart} shows an example process of how a human agent could decide which practice it should perform in a simulated scenario. For this illustration, it is deliberately held simple to leave room for discussion and to point towards open questions. First, the agent updates its interpretation of the context. At this point it is already important which practices it sees in its surroundings, because the context restricts the set of practices that can be performed (step 2). For each practice, the agent checks whether they disturb surrounding practices (as in the previous example, where the music practice disturbed the reading practice), and whether it would be disturbed by the surrounding practices. Only if no practice is disturbed, the selected practice will be performed.

We would like to point out a few characteristics of this simple model. First, it includes a strict yes/no decision process, which could be more rigid than what can be observed in the real world. One possibility would be to extend it with a probabilistic function to (rarely) allow for inappropriate practices. Second, the process provides no information about which ``surrounding'' practices are considered. A network model would probably consider practices that are performed by agents with direct links, but ABM also allows for spatial representations and dynamic models in which agents move around. Third, the list of possible practices that fit with the context is not included, and an overview of contexts that can be selected is also missing. And fourth, it is not clear from the flowchart which data are collected. One could count the performances of different practices, how often a specific practice is discarded because it disturbs or is disturbed by a surrounding practice, as well as the individual assessments of the context.

\section{Outlook}

We hope that our simple example illustrates the previous argument that ABMs spark reflection on several small decisions that have to be made, some way or another, in the real world. They help putting oneself in the agent's position and think through the process from its subjective perspective. This already provides some heuristic value and helps to point towards missing pieces in the theory. A functioning model furthermore points out weaknesses that are not obvious when considering only one agent, and helps to visualize and experiment.

Our longer-term ambition concerning ABMs and social acceptability is to find a manageable ``middle ground'' of sensitivity for social context that proves useful for designers. Currently, designers can choose between two groups of approaches when considering context. The first one is to not consider social context at all, or only to a very limited extent. Over time and through trial and error, changes to the design can then be implemented as reactions to the most important context restrictions to provide solutions that are considered as just acceptable. The second approach is to deeply engage with a specific context and adapt the design to it (e.g.,~\cite{Kuutti2014}). Further iterations can inductively make the design fit with other contexts. We think that both approaches could profit from a more informed understanding of context that helps focus on the important aspects, and a more manageable but rich understanding of how ``fit'' for different contexts is determined could improve the design process.

%overlapping contexts (homeoffice)

% How do people make assessments about what context they are in and what is allowed or not?

% Aus dem LBW: "description and analysis of interaction between different practices, methods have not been developed yet [in HCI]."

% We think that ABMs can be useful for that:
% - they allow for models including people and practices as active agents
% - they are dynamic
% - they have been used before with social practice theory
% 1. Holtz with a focus on the evolution of social practices over time~\cite{Holtz2014}
% 2. X with a focus on practice complexes of energy consumption in households
% 3. Compatability as in social acceptability has not been studied yet.
% - Other: Allow to experiment (group size, introducing unacceptable practices); study global patterns based on local interactions, allow for rich environments, study tipping points, patterns, path dependence, forces to make explicit

% Social Practice theory has received growing attention in HCI

% For individual practices, DfW and friends

%%%%%%%%%%%%%%%%%%%%%%%%

%Guideline Questions:
% Who are the central actors?
% What are the processes that happen between the different actors?
% What are the patterns that emerge?
% "Distributed multi-tasking"
% divergence between controlled and field

%%
%% The acknowledgments section is defined using the "acks" environment
%% (and NOT an unnumbered section). This ensures the proper
%% identification of the section in the article metadata, and the
%% consistent spelling of the heading.
\begin{acks}
This project is funded by the \grantsponsor{501100001659}{Deutsche Forschungsgemeinschaft (DFG, German Research Foundation)}{https://doi.org/10.13039/501100001659} – Grant No.~\grantnum{425827565}{425827565} and is part of~\grantnum{427133456}{Priority Program SPP2199 Scalable Interaction Paradigms for Pervasive Computing Environments}.
\end{acks}

%%
%% The next two lines define the bibliography style to be used, and
%% the bibliography file.
\bibliographystyle{ACM-Reference-Format}
\balance
\bibliography{/home/alarith/Dokumente/bibliography}

\end{document}